\documentclass[aps,letterpaper,twocolumn,prl,showpacs]{revtex4-1}
\usepackage{amssymb}
\usepackage{amsmath,bm}

\usepackage[pdftex]{graphicx,color}
\usepackage[english]{babel}
\usepackage{hyperref}
\usepackage{epstopdf}

\begin{document}

\title{Spontaneous symmetry breaking and trapping of temporal Kerr cavity solitons by pulsed or amplitude modulated  driving fields}

\author{Ian Hendry$^1$}
\email{ihen934@aucklanduni.ac.nz}
\author{Wei Chen$^{1,2}$}
\author{Yadong Wang$^1$}
\author{Bruno Garbin$^1$}
\author{Julien Javaloyes$^3$}
\author{Gian-Luca Oppo$^4$}
\author{St\'ephane Coen$^1$}
\author{Stuart G. Murdoch$^1$}
\author{Miro Erkintalo$^1$}
\email{m.erkintalo@auckland.ac.nz}

\affiliation{\vspace{10pt} $^1$The Dodd-Walls Centre for Photonic and Quantum Technologies, Department of Physics, The University of Auckland, Auckland 1142, New Zealand}
\affiliation{$^2$College of Meteorology and Oceanology, National University of Defense Technology, Changsha 410073, China}
\affiliation{$^3$Departament de F\'isica, Universitat de les Illes Balears, C/Valldemossa km 7.5, 07122 Mallorca, Spain}
\affiliation{$^4$SUPA and Department of Physics, University of Strathclyde, Glasgow G4 0NG, Scotland, EU}

\begin{abstract}
We report on a systematic study of temporal Kerr cavity soliton dynamics in the presence of pulsed or amplitude modulated driving fields. In stark contrast to the more extensively studied case of \emph{phase} modulations, we find that Kerr cavity solitons are not always attracted to maxima or minima of driving field amplitude inhomogeneities. Instead, we find that the solitons are attracted to temporal positions associated with specific driving field values that depend only on the cavity detuning. We describe our findings in light of a spontaneous symmetry breaking instability that physically ensues from a competition between coherent driving and nonlinear propagation effects. In addition to identifying a new type of Kerr cavity soliton behaviour, our results provide valuable insights to practical cavity configurations employing pulsed or amplitude modulated driving fields.
\end{abstract}
\maketitle

Temporal Kerr cavity solitons (CSs) are pulses of light that can recirculate indefinitely in coherently-driven, dispersive, Kerr nonlinear resonators~\cite{wabnitz_suppression_1993}. They were first observed in 2010~\cite{leo_temporal_2010}, and have attracted considerable attention ever since. Initial studies focused on macroscopic fiber ring resonators~\cite{leo_temporal_2010, leo_dynamics_2013, jang_ultraweak_2013, jang_observation_2014, jang_temporal_2015, luo_spontaneous_2015, anderson_observations_2016}, and were motivated by the prospect of using CSs as bits in all-optical buffers~\cite{jang_all-optical_2016}. In 2014, temporal CSs were also observed in high-Q nonlinear microresonators~\cite{herr_temporal_2014}, and they are now recognized to underlie the coherent and broadband ``Kerr'' optical frequency combs generated in such devices~\cite{coen_modeling_2013, chembo_spatiotemporal_2013, parra-rivas_dynamics_2014, yi_soliton_2015, brasch_photonic_2016, joshi_thermally_2016, webb_experimental_2016, marin-palomo_microresonator-based_2017, yi_single-mode_2017, weiner_frequency_2017, brasch_self-referenced_2017, pfeiffer_octave-spanning_2017, xue_soliton_2017, pasquazi_micro-combs:_2017}.

Temporal CSs are phenomenologically akin to spatial localized structures~\cite{ackemann_chapter_2009} that have been extensively studied in diffractive resonators~\cite{barland_cavity_2002}. In particular, similarly to their spatial counterparts, temporal CSs exhibit a property known as ``plasticity''~\cite{barland_temporal_2017}: inhomogeneities in the quasi-continuous background on top of which the CSs sit can cause motion along the dimension of localization. In the presence of perturbations that excite narrowband resonances in the soliton spectrum, such inhomogeneities can be generated by the solitons themselves, which can result in the formation of robustly bound soliton states and soliton crystals~\cite{wang_universal_2017,parra-rivas_interaction_2017,cole_soliton_2017}. Inhomogeneities can also be externally induced by shaping the quasi-continuous wave laser driving the cavity~\cite{firth_optical_1996, maggipinto_cavity_2000}. In particular, phase modulation of the cavity driving field has been shown to cause solitons to drift with a rate proportional to the local phase gradient~\cite{firth_optical_1996}, permitting robust trapping of CSs at the phase maxima~\cite{pedaci_all-optical_2008,pedaci_positioning_2006,jang_temporal_2015, jang_controlled_2016}.

The physics of (temporal) Kerr CSs in the presence of driving field \emph{phase} inhomogeneities has been thoroughly studied and is well understood. Although numerous studies have also investigated the behaviour of Kerr cavities in the presence of pulsed or amplitude modulated driving fields~\cite{steinmeyer_dynamical_1995,mitschke_soliton_1998,  garcia-mateos_optical_1995, torres_bilateral_1999, xu_experimental_2014, schmidberger_multistability_2014, parra-rivas_effects_2014, hansson_bichromatically_2014, cardoso_localized_2017, malinowski_optical_2017,xue_soliton_2017}, the dynamics of CSs and their trapping in such configurations has not yet been extensively examined. Early theoretical studies~\cite{maggipinto_cavity_2000}, based on the celebrated Lugiato-Lefever equation (LLE)~\cite{lugiato_spatial_1987}, have predicted dynamics similar to phase modulation, i.e., CSs moving along amplitude gradients towards maxima of the driving field. Yet, recent studies have shown anecdotal evidence of altogether different behaviours. Anderson et al. have found that a CS sitting atop an amplitude modulated background can be trapped at the edge of the modulation~\cite{anderson_coexistence_2017}. Similarly, Obrzud et al. have found, when simulating the generation of soliton frequency combs in microresonators driven with optical pulses, that the intracavity CSs can be temporally offset from the driving pulse center~\cite{obrzud_temporal_2017}.

On the one hand, the findings cited above~\cite{anderson_coexistence_2017, obrzud_temporal_2017} are somewhat surprising in light of the systems' parity symmetry. Indeed, localized structures are intuitively expected to be attracted towards non-zero parameter gradients only in systems with broken parity symmetry (e.g. in the presence of convection)~\cite{parra-rivas_effects_2014, javaloyes_dynamics_2016, camelin_electrical_2016}. On the other hand, Kerr cavities are well-known to exhibit spontaneous symmetry breaking~\cite{garcia-mateos_optical_1995, torres_bilateral_1999, xu_experimental_2014, schmidberger_multistability_2014,rossi_spontaneous_2016, cao_experimental_2017, bino_symmetry_2017}, which could explain the emergence of asymmetric states consisting of CSs trapped at edges of amplitude modulations. As a matter of fact, it has been explicitly noted that the profiles emerging from such symmetry breaking can bear some resemblance to CSs~\cite{xu_experimental_2014}. Moreover, prior studies have shown that localized structures of parity symmetric nonlinear systems (other than the LLE) can be attracted to positions offset from the perturbation extrema~\cite{scroggie_reversible_2005, fedorov_effects_2001}. For example, Scroggie et al. have shown this type of behaviour to arise almost universally when the perturbation varies rapidly (or comparably) compared to the width of the localized structure~\cite{scroggie_reversible_2005}. For the more specific case of slowly-varying amplitude modulations studied in this work, solitons drifting away from the modulation maximum has been noted in the context of the Swift-Hohenberg equation~\cite{scroggie_reversible_2005} as well as in the context of quadratically nonlinear optical resonators~\cite{fedorov_effects_2001}.  However, for Kerr CSs, similar behaviours have not yet been fully described or studied. Because of the growing interest in Kerr cavity configurations employing pulsed or amplitude modulated driving fields~\cite{obrzud_temporal_2017, obrzud_microphotonic_2017}, there is a need to gain better understanding of the behaviour of CSs in such systems.

In this Article, we report on a numerical study of Kerr CS dynamics in the presence of driving fields with inhomogeneous amplitude profiles. We find that, similarly to the case of quadratically nonlinear resonators~\cite{fedorov_effects_2001}, CSs in Kerr resonators are \emph{not} in general attracted to amplitude maxima or minima of the driving field. Instead, we find that the solitons are attracted to positions associated with a specific driving field amplitude whose value depends on the cavity detuning but is \emph{independent} of the local gradient. By identifying this new type of Kerr CS behaviour, our results could have impact on practical systems relying on pulsed or amplitude modulated driving fields, such as synchronously-driven microresonators~\cite{obrzud_temporal_2017, obrzud_microphotonic_2017} and fiber ring resonators~\cite{anderson_observations_2016, anderson_coexistence_2017}.

We consider a dispersive, Kerr-nonlinear ring resonator that is driven with a train of pulses or an amplitude modulated continuous wave field. We assume that the periodicity of the driving field is synchronized with the cavity round trip time, and that the resonator exhibits anomalous dispersion. The evolution of the slowly-varying intracavity field envelope $E(t,\tau)$ is then described by the following dimensionless mean-field LLE~\cite{haelterman_dissipative_1992, coen_modeling_2013, chembo_spatiotemporal_2013, xu_experimental_2014, parra-rivas_effects_2014}:
\begin{equation}
  \label{LLN0}
  \frac{\partial E(t,\tau)}{\partial t} = \left[ -1 +i(|E|^2- \Delta)
   +i\frac{\partial^2}{\partial\tau^2}\right]E+S(\tau).
\end{equation}
Here, $t$ is a slow time variable that describes the evolution of the slowly-varying intracavity field envelope $E(t,\tau)$ at the scale of the cavity photon lifetime, while $\tau$ is a corresponding fast time that describes the envelope's temporal profile over a single round trip. The terms on the right-hand side of Eq.~\eqref{LLN0} describe, respectively, the cavity losses, the Kerr nonlinearity, the cavity phase detuning, the group-velocity dispersion, and the (fast) time dependent coherent driving. Our normalization is the same as in ref.~\cite{leo_temporal_2010}: ${t \rightarrow \alpha t/t_\mathrm{R}}$, ${\tau \rightarrow \tau[2\alpha/(|\beta_2|L)]^{1/2}}$, and ${E \rightarrow E[\gamma L/\alpha]^{1/2}}$. Here $t_\mathrm{R}$ is the cavity roundtrip time, $\alpha$ is equal to half the fraction of power lost per round trip, $L$ is the resonator length, $\beta_2 < 0$ is the group-velocity dispersion coefficient, and $\gamma$ is the Kerr nonlinearity coefficient. The normalized cavity detuning $\Delta = \delta_0/\alpha$, where $\delta_0$ is the phase detuning of the pump from the closest cavity resonance. Finally, the normalized driving field amplitude $S(\tau) = E_\mathrm{in}(\tau)[\gamma L \theta/\alpha^3]^{1/2}$, where $E_\mathrm{in}(\tau)$ is the amplitude of the electric field injected into the resonator with units of $\sqrt{W}$, and $\theta$ is the input coupler power transmission coefficient. We note that, because we are assuming the driving field to be synchronous with the cavity round trip time, $S(\tau)$ does not depend on the slow time $t$ and Eq.~\eqref{LLN0} does not contain any convective drift terms~\cite{parra-rivas_effects_2014, javaloyes_dynamics_2016, coen_convection_1999}.

In all the calculations that will follow, we assume the driving field amplitude $S(\tau)$ to vary slowly in comparison to the CS duration.  In this case, the solitons experience a quasi-homogeneous driving (and hence can exist) but are perturbed by the underlying (quasi-linear) amplitude gradient~\cite{maggipinto_cavity_2000}. Dynamics in the presence of more rapidly-varying driving fields, which have been shown to give rise to ``reversible'' soliton motion in other nonlinear systems~\cite{scroggie_reversible_2005}, is beyond the scope of our present work.

We begin by considering a situation where a Kerr resonator is driven by a train of Gaussian pulses separated by the cavity round trip time. In this case, the driving field assumes the form
\begin{equation}
\label{Gau}
S(\tau) = S_0 \, \text{exp}\left(-\frac{\tau^2}{2\tau_\mathrm{G}^2}\right),
\end{equation}
where the duration $\tau_\mathrm{G} = 20$ is chosen to be much larger than the characteristic CS width ($\tau_\mathrm{CS} < 1$). To study the CS dynamics, we numerically integrate Eq.~\eqref{LLN0} with an initial condition that comprises of a short perturbation approximating a CS~\cite{wabnitz_suppression_1993, coen_universal_2013} that is offset from the driving field maximum: $E(0,\tau) = \sqrt{2\Delta}\,\text{sech}[\sqrt{\Delta}(\tau-\tau_0)]$. This perturbation reshapes into a CS which may (or may not) drift due to the amplitude gradient of the driving field. We run the simulation until steady-state -- where the CS no longer drifts -- is reached.

\begin{figure}[b]	
		\includegraphics[width=\linewidth]{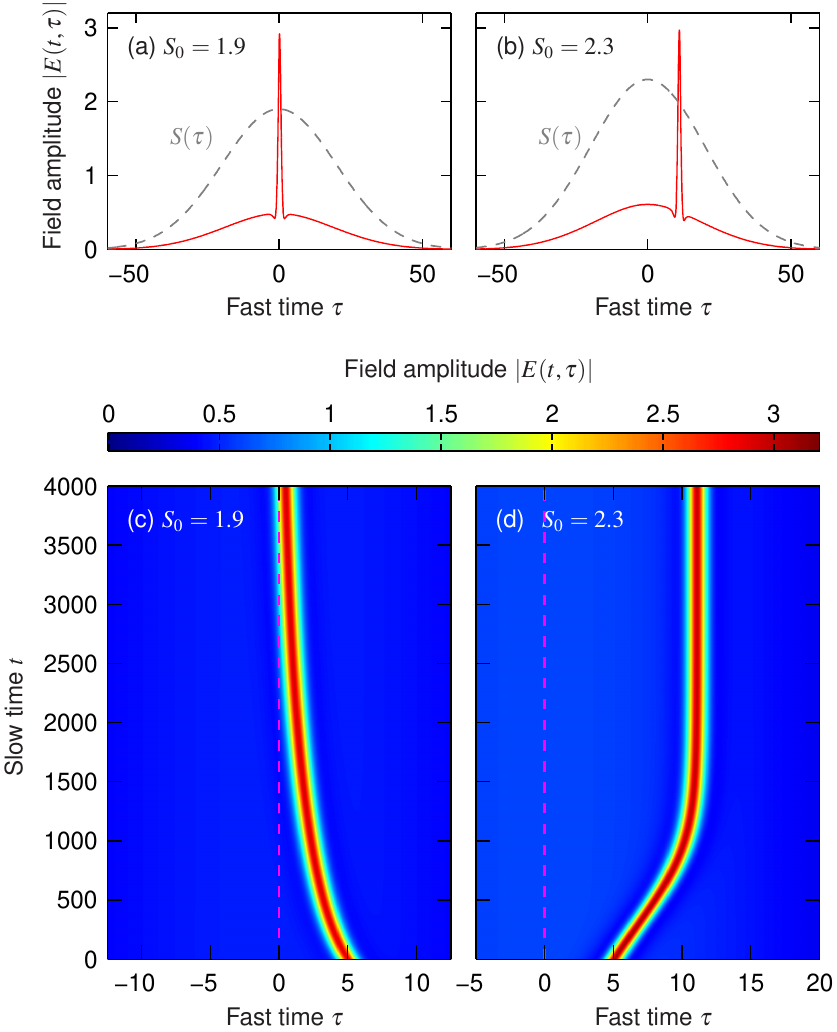}
		\caption{(a,b) Steady-state intracavity field solutions (red curves) for peak driving amplitudes (a) $S_0 = 1.9$ and (b) $S_0 = 2.3$. Gray dashed curves show the corresponding Gaussian driving field profiles. (c,d) Dynamical intracavity field evolutions corresponding to (a) and (b), respectively. The initial soliton position $\tau_0 = 5$. Dashed vertical magenta line highlights the position of the maximum driving field amplitude ($\tau = 0$). Note the different $x$-axes in (c) and (d).}
		\label{Fig1}
\end{figure}

Figures~\ref{Fig1}(a) and (b) show steady-state field profiles obtained for a constant detuning $\Delta = 4$ but for two different driving field amplitudes $S_0 = 1.9$ and $S_0 = 2.3$, respectively. Corresponding dynamical evolutions of the intracavity fields are shown as the false colour plots in Figs~\ref{Fig1}(c) and (d). As can be seen, for $S_0 = 1.9$ the CS is attracted to the peak of the driving field amplitude at $\tau = 0$. In contrast, for $S_0 = 2.3$, the CS drifts down along the driving pulse profile, eventually stabilizing at $\tau_\mathrm{CS} \approx 11.0$ where $S(\tau_\mathrm{CS})\approx 1.98$. Additional simulations (not shown here) reveal that, if the CS is initially excited slightly below the observed trapping point ($\tau_0>\tau_\mathrm{CS}$), it will move up along the driving pulse profile until it again stabilizes at $\tau_\mathrm{CS} \approx 11.0$.  Moreover, if the soliton is initially excited at $\tau_0 < 0$, it will be attracted towards $\tau_\mathrm{CS} \approx -11.0$ where $S(\tau_\mathrm{CS})\approx 1.98$. This clearly shows that, as expected based on symmetry considerations, the sign of the intensity gradient of the driving field plays no role in determining the CS's equilibrium position.

\begin{figure}[b]	
		\includegraphics[width=\linewidth]{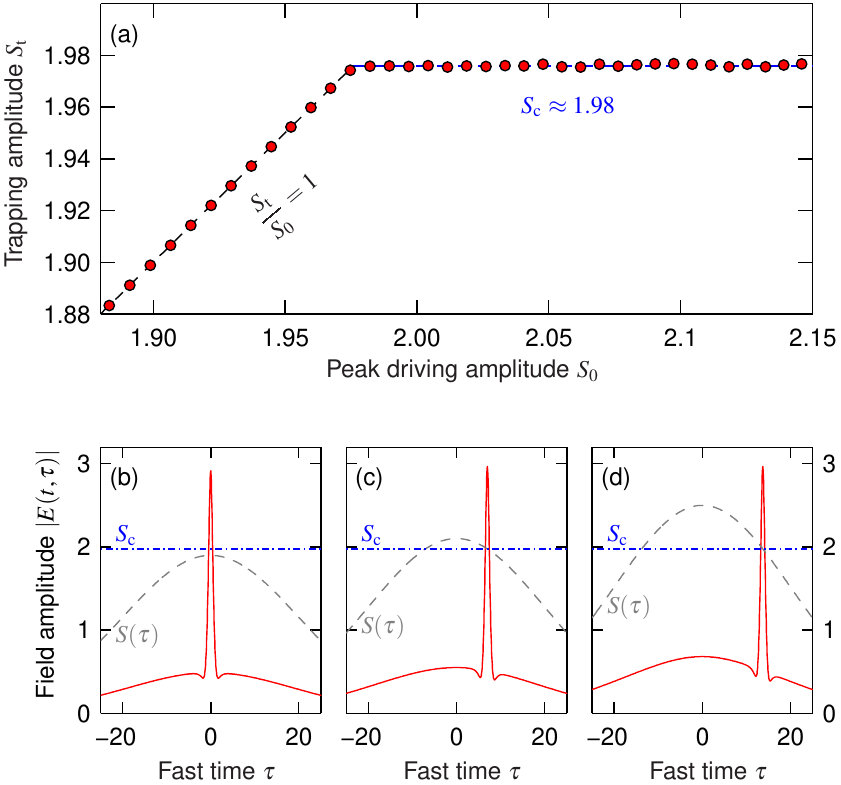}
		\caption{(a) Driving field amplitude at the steady-state CS position [$S_\mathrm{t} = S(\tau_\mathrm{CS})$] as a function of the peak driving amplitude $S_0$. (b)--(d) Red curves show steady-state intracavity field profiles for three different driving amplitudes: (b)~$S_0 = 1.9$, (c) $S_0 = 2.1$, and (d) $S_0 = 2.5$. Gray dashed curves show the corresponding Gaussian driving field profiles $S(\tau)$. Dash-dotted horizontal blue line indicates the critical driving value $S_\mathrm{c} = 1.98$. A detuning $\Delta = 4$ was used in all calculations.
}
		\label{Fig2}
\end{figure}

\begin{figure}[b]	
		\includegraphics[width=\linewidth]{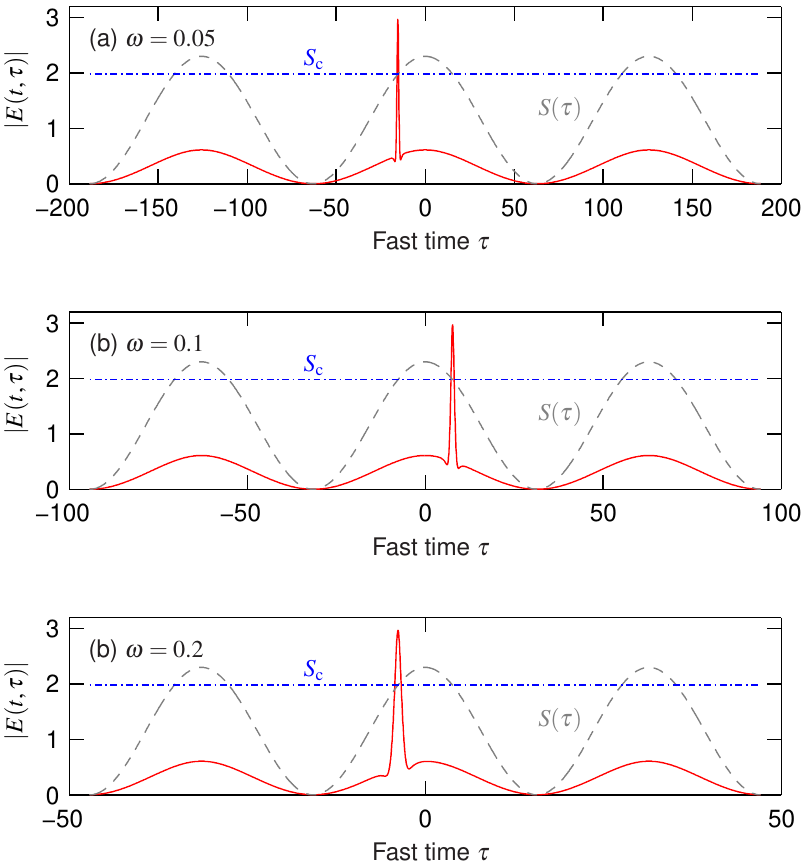}
		\caption{Steady-state field profiles (red curves) for a co-sinusoidal driving field with different modulation frequencies as indicated. The CSs were initially excited at (a) $\tau_0 = -1$, (b) $\tau_0 = 1$, and (c) $\tau_0 = -1$. Gray dashed curves show the corresponding driving field profiles, while the dash-dotted horizontal blue curve highlights $S_\mathrm{c} \approx 1.98$. All calculations use $S_0 = 2.3$ and $\Delta = 4$. Note the different fast time axes in (a)--(c).
}
		\label{Fig3}
\end{figure}

To gain more insights, we repeated the simulations above for a range of driving pulse amplitudes $S_0$. For each simulation, we extracted the final position $\tau_\mathrm{CS}$ of the CS relative to the peak of the driving pulse (at $\tau = 0$), as well as the corresponding value of the driving field at this point, i.e., $S_\mathrm{t} = S(\tau_\mathrm{CS})$. In Fig.~\ref{Fig2}, we plot $S_\mathrm{t}$ as a function of the peak driving amplitude $S_0$. For small $S_0$, we find $S_\mathrm{t} = S_0$: the CS is attracted and trapped to the peak of the driving field. However, when the peak driving amplitude increases beyond a critical value $S_\mathrm{c} \approx 1.98$, the soliton is always found to drift to a position with that driving field value. This behaviour is illustrated in Figs~\ref{Fig2}(b)--(d), where we plot the steady-state field profiles corresponding to three different driving peak amplitudes (see caption). In Fig.~\ref{Fig2}(b), the peak driving amplitude $S_0<S_\mathrm{c}$, and so the CS is attracted to the maximum of the driving profile. In contrast, in Figs.~\ref{Fig2}(c) and (d), the driving field encompasses the critical value $S_\mathrm{c}$, causing the CS to be trapped at one of the positions where $S(\tau_\mathrm{CS}) = S_\mathrm{c}$. We again emphasize that, depending on the initial condition, the CS can be trapped on either side of the Gaussian where $S(\tau) = S_\mathrm{c}$.

The results above suggest that Kerr CSs are attracted to positions where the driving field attains the critical value $S_\mathrm{c}$. It is only when the maximum driving amplitude is less than that critical value [as in Fig.~\ref{Fig2}(b)] that the soliton stabilizes at a maximum of the driving field. (Conversely, if the \emph{minimum} of the driving amplitude is larger than the critical value, the soliton will be trapped at the minimum of the driving field.) Similar behaviour has previously been attributed to CSs of quadratically nonlinear resonators~\cite{fedorov_effects_2001}. Through extensive simulations, we have found that the critical trapping level $S_\mathrm{c}$ does not depend on the driving field profile or on the amplitude gradient at the trapping point. To illustrate this, we consider a sinusoidally amplitude modulated driving field of the form
\begin{equation}
S(\tau) = \frac{S_0}{2}\left[1+\cos(\omega \tau)\right].
\label{sine}
\end{equation}
Figures~\ref{Fig3}(a)--(c) show steady-state profiles for three different modulation frequencies $\omega$ (see caption) with the peak driving amplitude and detuning held constant [at $S_0 = 2.3$ and $\Delta = 4$ as in Fig.~\ref{Fig1}(b)]. Here, to illustrate how the solitons can be trapped both at the rising or the falling edge of the driving field, different initial positions $\tau_0$ were used [see caption]. Despite the differences in driving field gradients ($S'(\tau_\mathrm{CS})\propto\omega$), in each case the CS is found to trap at a position where $S(\tau_\mathrm{CS}) \approx 1.98$, which coincides with the value found for pulsed driving in Fig.~\ref{Fig1}(b).

\begin{figure}[t]	
		\includegraphics[width=\linewidth]{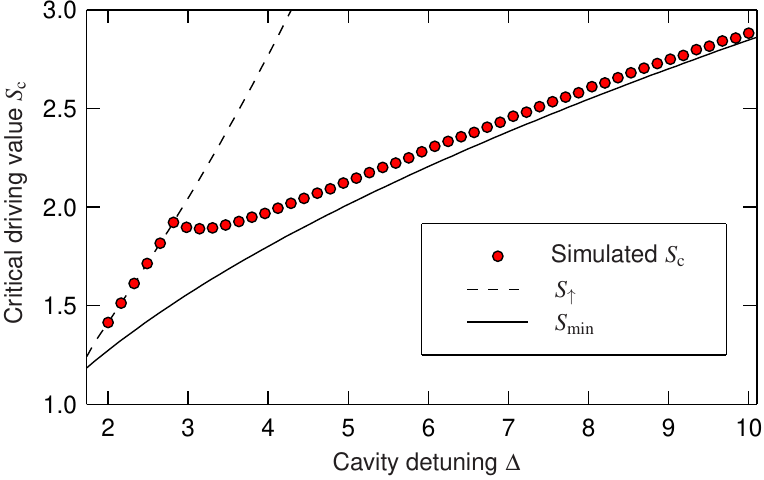}
		\caption{Critical driving field values $S_\mathrm{c}$ as a function of detuning. Red solid circles correspond to values extracted from numerical simulations, while solid black curve highlights the minimum driving amplitude needed for CS existence: $S_\mathrm{min} = (8\Delta/\pi^2)^{1/2}$. Dashed black curve highlights the up-switching point $S_\uparrow = [2/27(\Delta^3+9\Delta + \sqrt{\Delta^2-3})]^{1/2}$, above which the homogeneous response of the LLE is monostable. CSs can exist between the dashed and solid curves.}
		\label{Fig4}
\end{figure}

Whilst the critical driving amplitude $S_\mathrm{c}$ does not depend on the overall driving field profile, it does depend on the cavity detuning $\Delta$. We repeated the simulations above for a range of detunings, and extracted the driving field value towards which the CSs are attracted to. To ensure that the peak driving amplitude $S_0$ is sufficiently large to capture the critical value $S_\mathrm{c}$, we used a Gaussian driving profile with amplitude $S_0 = S_\uparrow = [2/27(\Delta^3+9\Delta + \sqrt{\Delta^2-3})]^{1/2}$. (This amplitude corresponds to the upper limit of the homogeneous bistability cycle of Eq.~\eqref{LLN0}, and hence the absolute upper limit of CS existence~\cite{leo_temporal_2010}.) In Fig.~\ref{Fig4}, we plot the critical driving value $S_\mathrm{c}$ obtained from our simulations as a function of the cavity detuning $\Delta$. Also shown are the maximum ($S_\uparrow$, dashed line) and minimum ($S_\mathrm{min}=(8 \Delta / \pi^2)^{1/2}$, solid line) driving field amplitudes between which CSs can exist~\cite{leo_dynamics_2013, herr_temporal_2014, barashenkov_existence_1996}. Two different regimes showing qualitatively different behaviour can be identified. For small $\Delta \lesssim 2.9$, we find $S_\mathrm{c}\approx S_\uparrow$: in this regime, \emph{a CS will always be attracted towards the local maximum of the driving field ($S_0$)}. In contrast, for larger $\Delta$, the trapping level $S_\mathrm{c}$ approaches the \emph{minimum} driving field amplitude $S_\mathrm{min}$. In this regime, the CS can be trapped at the edge of the driving field profile, and in the limit of $\Delta\gg1$, drift to the lowest possible value of $S$ for which it can still exist. This latter behaviour is similar to the dynamics observed in quadratically nonlinear systems~\cite{fedorov_effects_2001}, where soliton motion was explained by their tendency to approach conditions of nonlinear resonance. Indeed, we find that, for a given detuning $\Delta$, our Kerr CSs reach their maximum amplitude and they are precisely in-phase with the driving field when $S \approx (8 \Delta / \pi^2)^{1/2} \approx S_\mathrm{min}$, thus evidencing the realization of resonance conditions.

\begin{figure}[t]	
		\includegraphics[width=\linewidth]{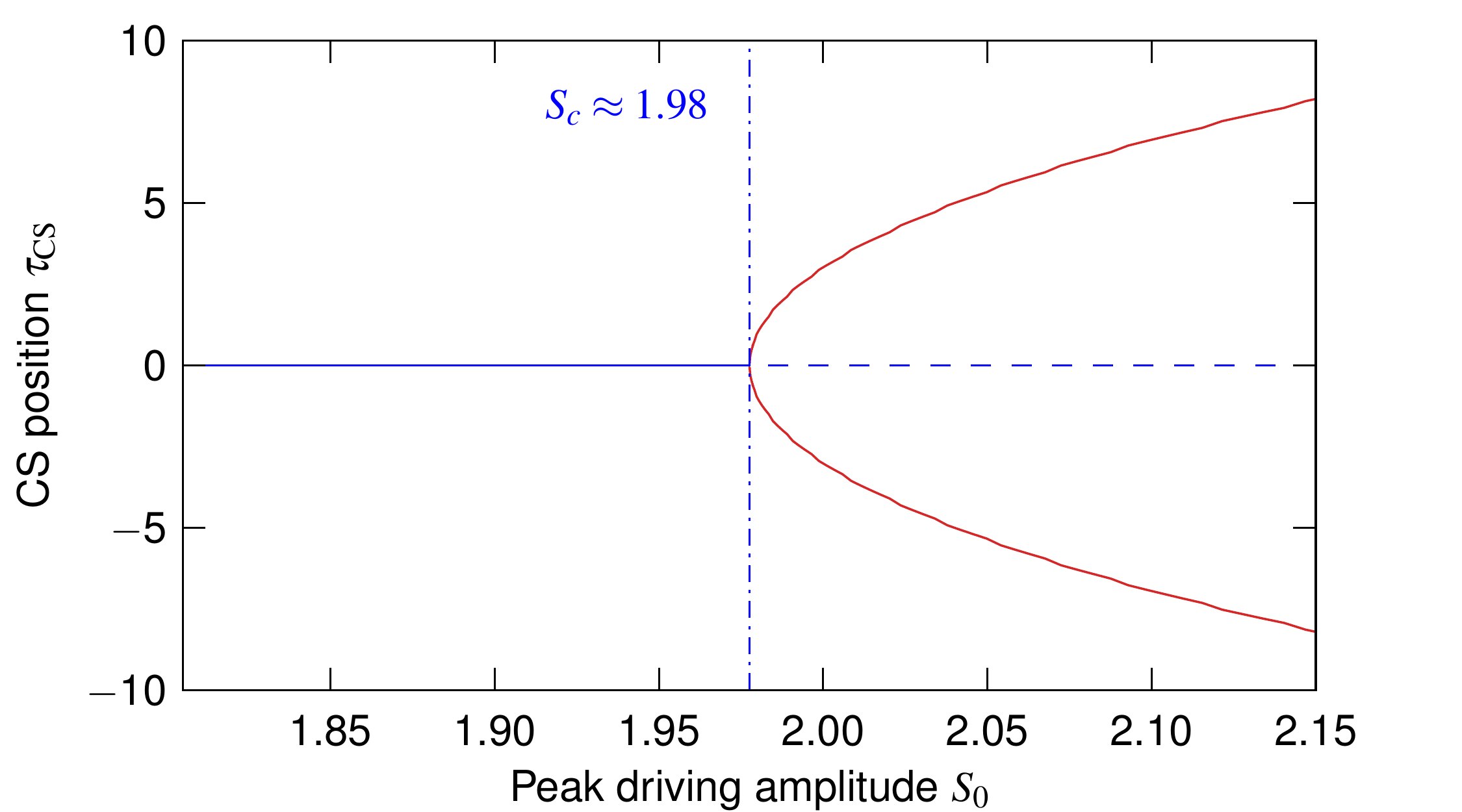}
		\caption{CS symmetry breaking bifurcation curve for a Gaussian driving field with $\tau_\mathrm{G} = 20$ and $\Delta = 4$. Blue and red curves show the steady-state positions of the CS solutions as a function of the maximum driving amplitude for symmetric and asymmetric states, respectively, with the dashed part being unstable. Dash-dotted vertical line indicates the critical driving value $S_\mathrm{c}$.}
		\label{Fig5}
\end{figure}

The observation that, for $\Delta\gtrsim 2.9$, Kerr CSs can be trapped at the edge of the driving field profile is amenable to an interpretation in terms of a spontaneous symmetry breaking instability~\cite{xu_experimental_2014}. This can be readily seen by plotting the possible steady-state CS positions, $\tau_\mathrm{CS}$, as a function of the peak driving strength $S_0$. An example of such a bifurcation curve is shown in Fig.~\ref{Fig5}; the steady-state field profiles were obtained using a Newton-Raphson continuation algorithm with a Gaussian driving field and $\Delta = 4$. For small $S_0$, the CSs sit stably atop the driving field maximum [c.f. Fig.~\ref{Fig2}(b)], and there is accordingly only a single steady-state configuration (with $\tau_\mathrm{CS} = 0$, blue curves). However, as $S_0$ increases past the critical level $S_\mathrm{c}$, a clear pitchfork bifurcation can be observed~\cite{Bruno}: the symmetric state with a CS at $\tau_\mathrm{CS} = 0$ becomes unstable, and a pair of new asymmetric stable states emerge that consist of a CS sitting on either side of the driving field maximum [red curves; see also Figs~\ref{Fig2}(c) and (d)].

It is somewhat surprising that Kerr CSs can be trapped at a position where the driving field amplitude gradient is non-zero. Indeed, to first order, the CS's drift velocity can be shown to be directly proportional to that gradient ~\cite{maggipinto_cavity_2000, javaloyes_dynamics_2016}:
\begin{equation}
v = \frac{d\tau_\mathrm{CS}}{dt} = a \left.\frac{dS}{d\tau}\right|_{\tau=\tau_\mathrm{CS}},
\label{driftv}
\end{equation}
where the proportionality coefficient $a$ describes the projection of the CS's neutral (or Goldstone) mode along a linear fast time variation. Note that amplitude modulations correspond to purely real perturbations and only couple to the real part of the neutral mode, in stark contrast to phase modulations. Accordingly, while Eq.~\eqref{driftv} holds true also for the case of phase modulated driving fields, the coefficient $a$ differs between the two forms of perturbations ~\cite{maggipinto_cavity_2000}.

\begin{figure}[t]	
		\includegraphics[width=\linewidth]{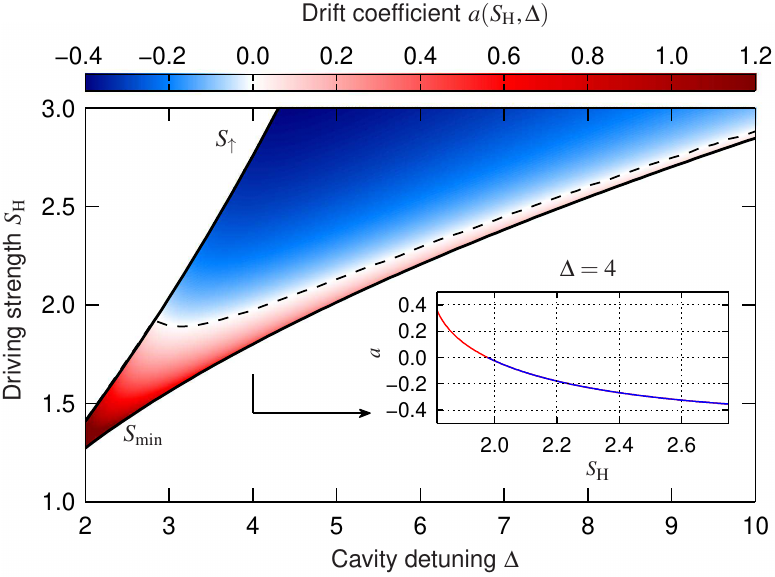}
		\caption{Drift coefficient $a(S_\mathrm{H},\Delta)$ calculated from the neutral mode of a steady-state CS solution for a range of homogeneous driving strengths $S_\mathrm{H}$ and cavity detunings $\Delta$. The solid black curves correspond to $S_\mathrm{min}$ and $S_\uparrow$ as defined in the caption of Fig.~\ref{Fig4}; CSs only exist between these two curves. Black dashed curve shows the critical driving field values $S_\mathrm{c}$ obtained from direct numerical simulations of the LLE with a Gaussian driving field [same data is shown in Fig.~\ref{Fig4}]. Inset shows the curve $a(S_\mathrm{H},\Delta = 4)$.}
		\label{Fig7}
\end{figure}

The apparent discrepancy between our findings and Eq.~\eqref{driftv} is explained by the fact that the CS's neutral mode changes with the driving strength (and detuning). As a consequence (and similarly to quadratically nonlinear systems~\cite{fedorov_effects_2001}), the proportionality coefficient $a$ in Eq.~\eqref{driftv} also depends on the driving strength (and detuning), i.e., $a = a(S_\mathrm{H},\Delta)$, where $S_\mathrm{H} = S(\tau_\mathrm{CS})$.  This is illustrated in Fig.~\ref{Fig7}, where we explicitly show $a(S_\mathrm{H},\Delta)$ computed for a range of cavity driving strengths and detunings~\cite{maggipinto_cavity_2000}. These results were obtained by first finding the steady-state CS solutions of Eq.~\eqref{LLN0} for a \emph{homogeneous} driving field with strength $S_\mathrm{H}$ and detuning $\Delta$, and then projecting the real part of the solitons' neutral mode (technically the odd components of the left eigenvector with zero eigenvalue of the system's Jacobian) along a linear fast time variation~\cite{maggipinto_cavity_2000}.

As can be seen, the coefficient $a$ decreases with increasing driving strength, and for $\Delta \gtrsim 2.9$, crosses zero within the region of CS existence. Moreover, we see that the curve $a(S_\mathrm{H},\Delta) = 0$ matches exactly with the critical driving field values found through direct split-step simulations of Eq.~\eqref{LLN0} with an inhomogeneous driving field [c.f. Fig.~\ref{Fig4}]. These findings fully corroborate our observations of CS behaviour in the presence of driving field amplitude inhomogeneities. Specifically, when a CS drifts along an amplitude gradient, the coefficient $a$ it experiences changes continuously. At the critical level $S_\mathrm{c}$, the coefficient passes through zero and changes sign, thus enabling robust trapping at that level. On the other hand, whilst the maximum (or minimum) of the driving field (with $dS/d\tau = 0$) always corresponds to an equilibrium position, that equilibrium position is unstable [c.f. Fig.~\ref{Fig5}] if the maximum (minimum) is larger (smaller) than the critical driving value $S_\mathrm{c}$. There is therefore no contradiction between our findings and Eq.~\eqref{driftv}: the soliton velocity is  ``locally'' proportional to the driving field gradient [as described by Eq.~\eqref{driftv}], but because the proportionality coefficient changes as the soliton drifts, the overall relationship is more complex. (Rigorously speaking, the soliton velocity is proportional to the driving field gradient only over short slow time intervals during which the ``local'' driving strength --- and hence the coefficient $a(S_\mathrm{H},\Delta$) --- experienced by the soliton remains approximately constant.) It is also worth highlighting that, because the drift coefficient $a(S_\mathrm{H},\Delta)$ only depends on the local value of the driving field ($S_\mathrm{H}$) and the detuning, the analysis above readily explains why the critical trapping level $S_\mathrm{c}$ does not depend on the precise profile of the driving field (provided that the driving field varies slowly compared to the CS duration).

To better understand the physics that underpins the Kerr CS behaviour identified above, we next present results from simulations of an Ikeda-like map~\cite{ikeda_multiple-valued_1979}. Unlike the mean-field approximation of Eq.~\eqref{LLN0}, this approach allows us to isolate effects due to (i) propagation through the Kerr medium over a single cavity round trip and (ii) the coherent injection of the driving field into the cavity. We write the map equations in dimensionless form with units that allow immediate comparison with results from Eq.~\eqref{LLN0}:
\begin{align}
\frac{\partial E_{m}(\xi,\tau)}{\partial \xi} &= i\frac{\partial^2 E_m}{\partial \tau^2 } +  i|E_m|^2 E_m, \label{NLSE} \\
E_{m+1}(\xi=0,\tau) &= \sqrt{1-2\alpha}\,E_{m}(\xi = \alpha,\tau)e^{-i\delta_0} + \alpha S(\tau). \label{drive}
\end{align}
Here, Eq.~\eqref{NLSE} is the well-known nonlinear Schr\"odinger equation (NLSE) that describes the evolution of the intracavity field over one cavity round trip, with $\xi = \alpha z/L$ a dimensionless propagation coordinate ($z$ is the corresponding dimensional variable), while Eq.~\eqref{drive} is the boundary condition that describes the addition of the coherent driving field to the intracavity light field at $\xi = 0$. For high-finesse cavities, $\alpha \ll 1$, and the above map equations can be averaged to the LLE given by Eq.~\eqref{LLN0}. To better capture the evolution of the soliton over one cavity round trip, we have used a comparatively large value of $\alpha = 0.15$ in the simulations that will follow.

\begin{figure}[htb]	
		\includegraphics[width=\linewidth]{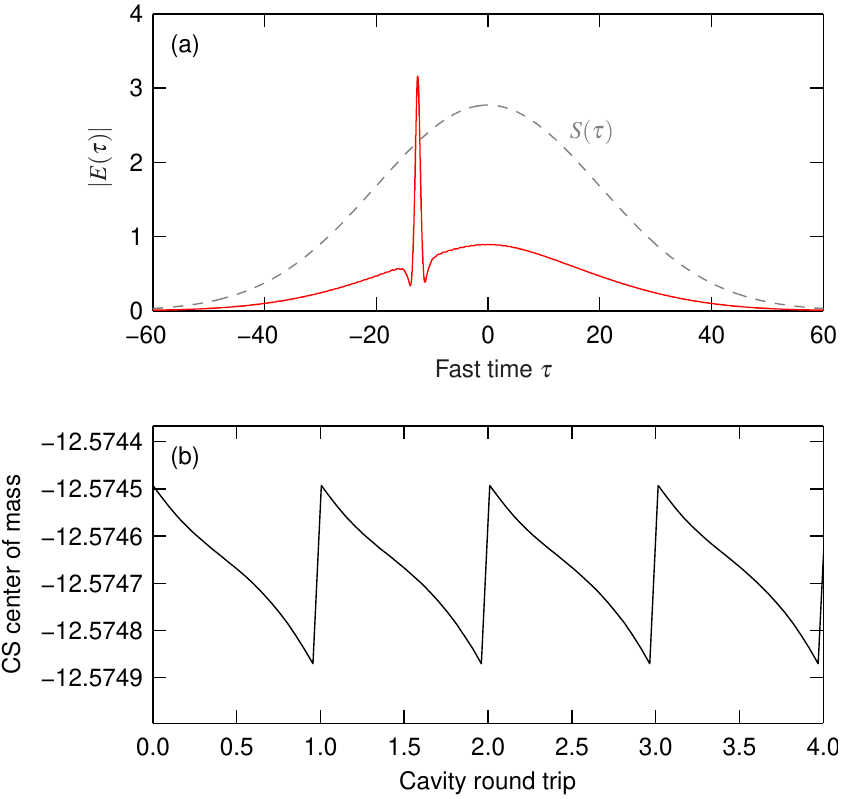}
		\caption{Simulation results from an Ikeda-like cavity map. (a) Steady-state intracavity field profile (red curve) for a Gaussian driving field (gray dashed curve). (b) Evolution of the CS's center of mass over four cavity transits.}
		\label{Fig6}
\end{figure}

Figure~\ref{Fig6}(a) shows a steady-state intracavity field obtained from the Ikeda map with $\Delta = 4$ and a Gaussian driving field profile with $S_0=S_\uparrow\approx2.77$ and $\tau_\mathrm{G} = 20$. One first notes that the Ikeda map reproduces the salient result of the LLE simulation, i.e., the CS trapping at a position where the driving field gradient is non-zero. The precise trapping value $S_\mathrm{c} \approx 2.27$ is somewhat larger than the value found in corresponding mean-field simulations, which we attribute to the comparatively large value of $\alpha$. Indeed, we have carefully verified that the Ikeda map reproduces the LLE result in the limit of very small $\alpha$.

To gain insights on the interplay between propagation over one round trip [described by Eq.~\eqref{NLSE}] and addition of the coherent driving [Eq.~\eqref{drive}], Fig.~\ref{Fig6}(b) shows the evolution of the CS's centre of mass in the fast time dimension (calculated over the soliton's half-maximum points) over four consecutive round trips after steady-state is reached. As can be seen, the CS drifts gently downwards away from the driving field maximum during propagation, but is pulled back to its original position at the boundary. This competition between propagation and coherent driving underpins the behaviour of CSs in the presence of pulsed or amplitude modulated driving fields. Specifically, if the propagation effect is stronger (weaker) than the driving effect, the coefficient $a$ in Eq.~\eqref{driftv} is negative (positive), such that the CS will drift away from (towards) the maximum. In contrast, at the critical driving strength $S_\mathrm{c}$, the two effects are precisely balanced.

The physics behind the two competing effects identified above can be qualitatively explained as follows. First, the addition of the driving field can be intuitively understood to shift the CS towards its maximum because the two are almost in-phase. In contrast, the soliton's drift away from the maximum during propagation is due to the phase shift between the soliton and the \emph{intracavity} background field. Considering a superposition field $E(\xi,\tau) = E_\mathrm{s}(\xi,\tau) + \delta E(\xi,\tau)$ that consists of an NLSE soliton ($E_\mathrm{s}$) perturbed by a small amplitude background field ($\delta E$), it is well known that the perturbation can cause the soliton to drift, with the rate of drift given by the inverse group velocity~\cite{gordon_random_1986}
\begin{equation}
\frac{\Delta\tau_\mathrm{S}}{\Delta\xi} = -\frac{1}{A}\mathrm{Im}\int\frac{\partial E_\mathrm{s}^*}{\partial\tau}\delta E\,\mathrm{d}t,
\label{gordoneq}
\end{equation}
where $\Delta\tau_\mathrm{S}$ and $A$ represent the soliton's temporal position and amplitude, respectively. Straightforward analysis of Eq.~\eqref{gordoneq} confirms that, when $\Delta\phi = \phi_\mathrm{S} - \phi_\mathrm{\delta} \in [0,\pi]$, where $\phi_\mathrm{S}$ and $\phi_\mathrm{\delta}$ denote respectively the phases of the soliton and the background, the soliton will drift away from the perturbation maximum while the opposite is true for $\Delta\phi > \pi$. (We have also confirmed these predictions by means of direct numerical simulations of the NLSE Eq.~\eqref{NLSE}.) For Kerr CSs, the two phases are approximately (in the mean-field limit) given by~\cite{wabnitz_suppression_1993, herr_temporal_2014}
\begin{align}
\phi_\mathrm{S} &\approx \cos^{-1}\left(\frac{\sqrt{8\Delta}}{\pi S}\right), \\
\phi_\mathrm{\delta} &\approx -\tan^{-1}\left(\Delta\right).
\end{align}
As CSs exist only for $\Delta > 0$, one always finds $\Delta\phi\in [0,\pi]$, explaining the soliton's downward motion over a single cavity round trip. It is worth noting that, if the soliton sits at an extremum of the driving field (where ${\partial E_\mathrm{s}^*/\partial\tau = 0}$), this motion vanishes [see Eq.~\ref{gordoneq}]. Because the addition of a parity symmetric driving field will likewise induce no shifts in this situation, we can see how the driving field extrema indeed correspond to equilibria, whose stability is governed by the relative strengths of the two competing effects.

To conclude, we have investigated the dynamics of Kerr CSs in the presence of driving fields with inhomogeneous amplitude profiles. In stark contrast to the case of phase inhomogeneities, we have shown that the CSs are not in general attracted to maxima (or minima) of an amplitude modulated driving field. Instead, the solitons are attracted to --- and trap to --- positions associated with particular values of the driving field. We have described the underlying physics in terms of a spontaneous symmetry breaking instability that arises from a competition between the coherent addition of the driving field and propagation in the Kerr medium.

Our work complements previous studies of symmetry breaking in Kerr cavities~\cite{garcia-mateos_optical_1995, torres_bilateral_1999,schmidberger_multistability_2014, xu_experimental_2014,rossi_spontaneous_2016,cao_experimental_2017, bino_symmetry_2017}, and raises several interesting questions for follow-up research: how do CSs behave in presence of amplitude \emph{and} phase inhomogeneities; does the universality of the general behavior evoked by amplitude inhomogeneities extend beyond Kerr cavities and the quadratically nonlinear systems studied in~\cite{fedorov_effects_2001}? Of course, experimentally verifying the predictions outlined in our current work also represents a significant future contribution. While some of us have already observed experimental evidence of CS trapping to the edge of a nanosecond pump pulse~\cite{wang_intensity_2017}, the results obtained do not allow unequivocal discrimination between effects arising from intrinsic cavity dynamics and non-ideal driving conditions (e.g. synchronization mismatch or residual pump phase modulation). On the other hand, clean experimental evidence of symmetry breaking has previously been observed in a fiber ring resonator driven with pulses from a mode-locked laser~\cite{xu_experimental_2014}. Although that study did not explore the connection between symmetry breaking and CS trapping, we believe that the experimental configuration used could allow for the controlled examination of the behaviours identified in our current work.

\begin{acknowledgments}
The authors wish to acknowledge the Centre for eResearch at the University of Auckland for their help in facilitating this research. We also acknowledge financial support from the Marsden Fund and the Rutherford Discovery Fellowships of the Royal Society of New Zealand.  J. J. additionally acknowledges funding from the project COMBINA (TEC2015-65212-C3-3-P AEI/FEDER UE), and W. C. acknowledges support from the National Natural Science Foundation of China (61505258).
\end{acknowledgments}

\end{document}